\documentclass[aps,pre,twocolumn,footinbib,longbibliography,superscriptaddress]{revtex4-2}
\usepackage{amsmath}
\usepackage{amsfonts}
\usepackage{amssymb}
\usepackage[dvipdfmx]{graphicx}

\begin{document}
\title{Ecosystem transformations in response to environmental fluctuations}

\author{Ikumi Kobayashi}
\affiliation{Department of Physics 1, Graduate School of Science, Kyoto University, Kyoto 606-8502, Japan}

\begin{abstract}
Ecosystems, which are intricate amalgams of biological communities and their surrounding environments, continually evolve under the influence of their myriad interactions. The world is currently facing intensifying environmental fluctuations. Understanding general trends in ecosystem transformations in response to environmental fluctuations and elucidating the underlying mechanisms are thus critical challenges. In this study, we used a model ecosystem approach to investigate ecosystem alterations caused by escalating environmental fluctuations. We analyzed two distinct models: a stochastic ecosystem model with a spatial structure, and a differential equation model for resource competition. We found that environmental fluctuations tend to shift multi-species coexistence toward the dominance of specific species. We also categorized biological species as specialists or generalists and discovered that which of these groups becomes the dominant species depends on the intensity and frequency of environmental fluctuations. We also determined that a qualitative change in the diversity-stability relationship depends on the period of environmental fluctuations. These results underscore the need to explicitly consider the type of perturbation when discussing ecological transitions and the stability of ecosystems. Our findings advance understanding of the mechanisms underlying how environmental changes reshape ecosystems and offer insights into ecosystem sustainability in the face of future environmental perturbations.
\end{abstract}

\date{\today}

\maketitle

Ecosystems are intricate systems comprising biological communities and their surrounding environments.
These systems are characterized by complex interplay among diverse species that fosters intricate networks of food chains, competitions, and symbiotic relationships. This web of interactions is fundamental for maintaining energy and nutrient flow, thereby preserving ecosystem balance. At the same time, environmental variables, such as temperature, precipitation patterns, and soil properties, exert profound influences on the organisms residing within these ecosystems and impose crucial constraints on ecosystem structure and functions. Moreover, ecosystem engineers, such as beavers, actively reshape their surroundings, thereby further enhancing the intricacies of these ecosystems \cite{jones1994organisms, wright2002ecosystem}. Ecosystem stability and sustainability thus depend on the intricate balance of these interactions.

As recently recognized, environmental change is intensifying in many respects, with consequent far-reaching effects \cite{aghakouchak2020climate, fischer2021increasing, walther2002ecological, bellard2012impacts, sintayehu2018impact, berry2010climate, stern2008economics}. One notable example is the trend of rising global temperatures, which has led to heightened instances of increasingly severe heatwaves in numerous regions \cite{rousi2022accelerated}. Alterations in precipitation patterns have induced prolonged droughts in specific locations \cite{dai2013increasing, buntgen2021recent} while concurrently triggering intensified rainfall and flood events elsewhere \cite{tabari2020climate}. Variations in sea surface temperatures have exerted substantial impacts on marine ecosystems, in turn reshaping the distribution and diversity of marine species \cite{chaudhary2021global}. In addition, anthropogenic activities, such as land-use changes and deforestation, have expedited habitat loss and fragmentation \cite{wilcox1985conservation, yasuda2000earliest, fahrig2019habitat}, thereby compounding the challenges confronting ecosystems \cite{ripple2017world}.

Given this context, clarifying the impacts of intensifying environmental fluctuations on ecosystems is important. Elucidation of these phenomena forms a linchpin for evaluating the sustainability of ecosystems and developing effective conservation strategies. Previous research on ecosystem risk assessment has relied on extensive global simulations, which combine climate model scenarios with dynamic global vegetation models \cite{scholze2006climate, gerten2004terrestrial, sitch2008evaluation, xu2020assessing}. Our approach, which entails analysis of simplified ecosystem models, provides a contrasting, complementary perspective. In addition to considering the responses of individual actors within ecosystems, recognition is growing that exploration of the complex nonlinear dynamics driven by interactions among ecosystem members is needed as well \cite{walther2010community, scheffer2003catastrophic, andersen2009ecological}. We have thus focused on pivotal ecological interactions, such as competition and predation, and examined the mechanisms through which these interactions either buffer or amplify the effects of environmental perturbations.

Our approach aligns with the longstanding tradition of mathematical and theoretical biology, which aims to capture the essence of the intricate dynamics of biological phenomena using simple models. The core aspiration of mathematical biology is to provide concise and insightful explanations for the rich tapestry of life processes. By distilling complex ecological and environmental interactions into mathematical frameworks, we aimed to uncover underlying principles and general trends that transcend specific case studies. We further note that these dynamics, which encompass phenomena such as global environmental fluctuations and species extinctions, are inherently challenging to explore through controlled field experiments. Integrating insights from real-world field observations with those derived from mathematical model analyses is a promising avenue to enhance our understanding and prediction of ecosystem transformations associated with environmental fluctuations.

To decipher the intricate responses of ecosystems to environmental fluctuations, we analyzed two distinct ecosystem models in this study.
The first model was a stochastic ecosystem model explicitly containing spatial structure, where organisms stochastically experience birth, death, migration, and speciation processes within a dynamic environment. The second model was a differential equation model for resource competition, where multiple species vie for multiple resources within a fluctuating environment. In addition, we categorized biological species as either specialists, adept at thriving within specific environmental conditions, or as generalists, adaptable to a wide range of conditions. Through these models, we explored the impacts of environmental changes on ecosystems, thereby shedding light on the underlying mechanisms governing these dynamic processes.

\section*{The Birth-Death-Diffusion Model}
We first describe a stochastic ecological model with a spatial structure. 
This model, a variation of Durrett and Levin's ecosystem model \cite{durrett1996spatial}, provides a simplified framework encompassing environmental variability, species’ adaptive strategies, and spatial patterns of communities. The model serves as the foundation for discussing the impacts of environmental fluctuations on ecosystems.

We consider a scenario where two distinct environmental conditions, $E = 1$ and $E = 2$, alternate periodically with a fixed duration $\tau$. These alternating conditions can be likened to the changing of seasons, such as the transition from summer to winter or from a dry season to a rainy one. These seasonal transitions introduce significant environmental variations that not only affect the availability of resources, but also cause physiological stress that influences species’ survival rates. Environmental stressors, such as summer heat waves, can range from moderate to extreme, thus making the level of stress quite diverse. To characterize the severity of these environmental stresses, we introduce the parameter $I$, which subsequently will be linked to the death rate of each species.

The biological communities are modeled on an $L\times L$ two-dimensional square lattice, where stochastic events of death, birth, and speciation take place. Each grid point is either occupied by some species $s$ or left blank. Under environment $E$, each individual of species $s$ experiences death at a rate of $d_s^E$. Birth and speciation events occur when at least one adjacent site is unoccupied. To simplify the model, we assume constant birth rates across both environments and species, denoting this as $b_s^E \equiv b_0$. Under environment $E$, species $s$ reproduces at a rate of $(1-\nu)b_0$ and fills a randomly chosen unoccupied neighboring site. In addition, species $s$ under environmental condition $E$ can undergo speciation at a rate of $\nu b_0$, which results in the emergence of a new species $s'$ at a randomly selected unoccupied neighboring site. Here, the parameter $\nu$ signifies the speciation probability.

Species are categorized as specialists or generalists on the basis of their adaptive strategies to their environments. We let $c_s^E$ be the species-specific physiological and ecological costs incurred for environmental adaptation. For each species $s$, the sum of these costs across all environments is normalized to 1, i.e., $\sum_{E=1,2} c_s^E = 1$. We assume death rates proportional to the environmental stress intensity and a decreasing function of the adaptation costs. As a simple functional form that satisfies these constraints, we adopt $d_s^E = I(1 - c_s^E)$. In this formulation, a high value of $c_s^E$ under a specific environmental condition signifies the species’ substantial investment in physiological and ecological adaptations to that particular environment, thus characterizing it as a specialist. Conversely, when a species allocates its resources relatively uniformly across various states, it is classified as a generalist, thereby highlighting its adaptability to a wide range of environmental conditions. Lastly, when speciation events occur, we randomly assign adaptation costs to the newly emerged species $s'$. Specifically, we set $[c_{s'}^1, c_{s'}^2] = [x, 1 - x]$, where $x$ is drawn as a uniform random variable from the interval $[0, 1]$.
For a more detailed description of the simulation procedure, please refer to the SI Appendix. In addition, the typical dynamics of this model are provided in SI Movie 1.

\section*{Results Obtained under  the Birth-Death-Diffusion Model}
\subsection*{Ecosystem Transformations}
\begin{figure}
\centering
\includegraphics[width=8.6cm]{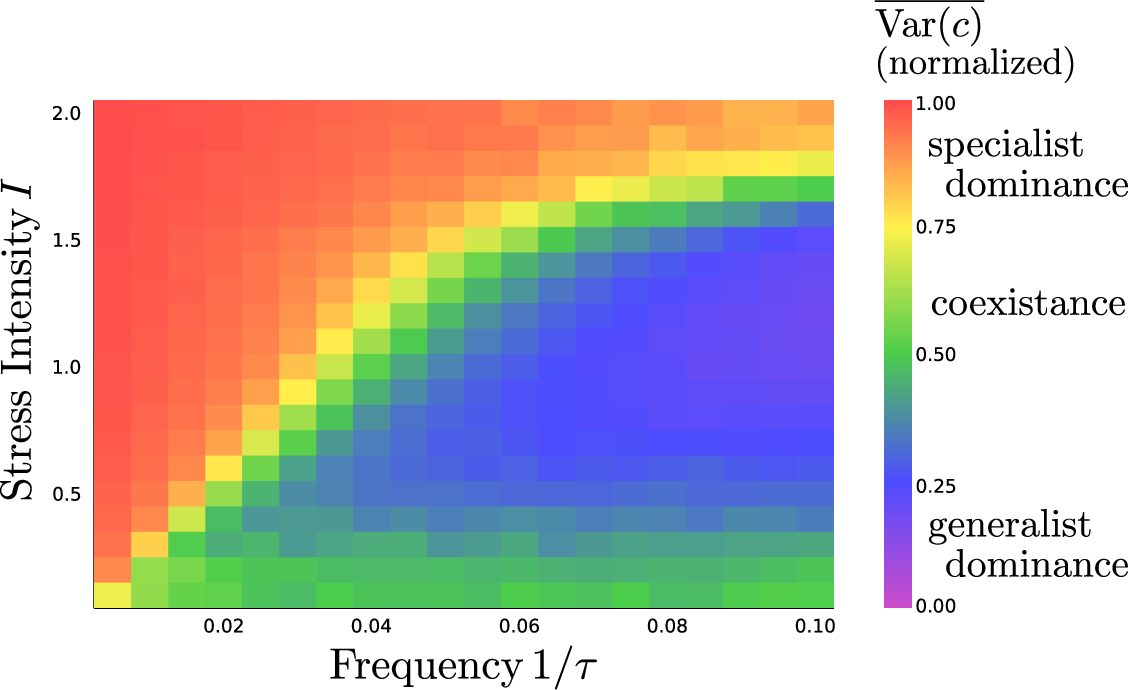}
\caption{
Prevalence of Specialist and Generalist Strategies.
This figure illustrates the prevalence of ecological strategies, either specialist or generalist, across different environmental conditions. We measured the long-time average of the variance of adaptation costs as an indicator of the degree of specialization. The parameter space is broadly categorized into three regions, each characterized by distinct domination patterns. Relatively slow environmental fluctuations tend to favor specialist dominance, whereas rapid fluctuations lead to the prevalence of generalists. Under conditions of weak environmental stress, specialists and generalists coexist.
}
\label{result1}
\end{figure}

We systematically explored how ecosystems are transformed by varying the period and stress intensity of environmental fluctuations. Figure \ref{result1} comprehensively represents the prevalence of either ecological strategy, i.e., specialist or generalist, under diverse environmental conditions. Using the parameter values of $L=20, b_0=1.0$, and $\nu=0.01$, we measured the long-time average of the variance of adaptation costs, $\overline{\text{Var}(c)}$, as an indicator of the degree of specialization. When $\overline{\text{Var}(c)}$ is close to zero, environmental adaptation costs are distributed evenly between the two environmental states ($E=1$ and $E=2$), thus indicating that generalists fill the field. Conversely, a relatively large value of $\overline{\text{Var}(c)}$ indicates that specialists dominate the ecosystem.

The parameter space can be broadly categorized into three distinct regions. When environmental fluctuations have a relatively long period, specialists tend to dominate the ecosystem. Conversely, rapid fluctuations lead to the dominance of generalists. Under weak environmental stress, both specialists and generalists can coexist, thereby representing a diversity of adaptation strategies. This observed trend is consistent with the results of previous studies using different models \cite{richmond2005role, travis2003climate}.

We next examined the effects of heightened environmental fluctuations, which correspond to an escalation of the environmental stress intensity, $I$, within our model. Even when an initial state permits the coexistence of specialists and generalists, intensification of environmental stress $I$ triggers a transformation within the ecosystem, which favors the dominance of specific species contingent upon fluctuation frequency.

\subsection*{Diversity-Stability Relationship}
\begin{figure}
\centering
\includegraphics[width=8.0cm]{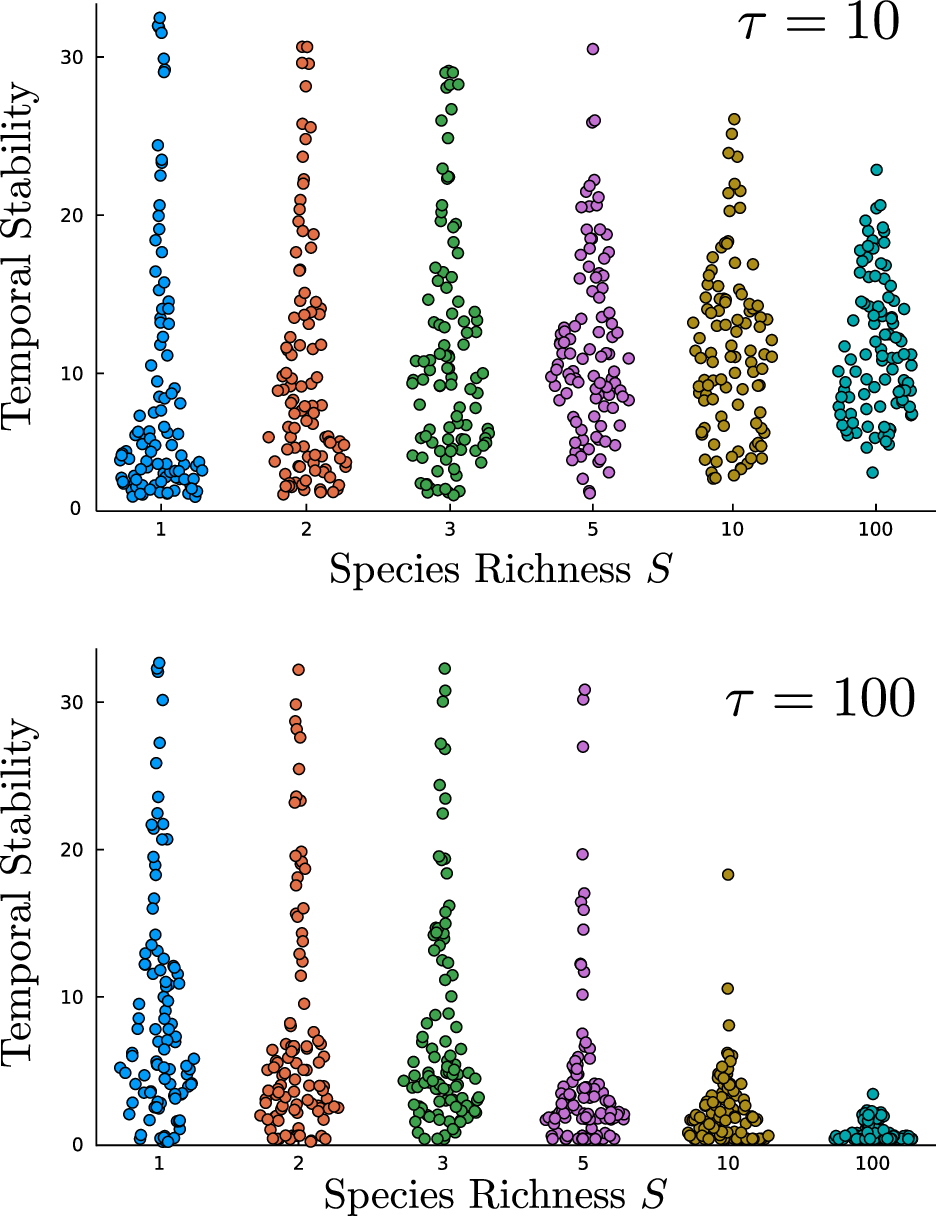}
\caption{
Diversity–Stability Relationship. 
This figure shows the relationship between species richness and temporal stability of total biomass under our model. The top graph corresponds to relatively rapid environmental fluctuations ($\tau=10$), whereas the bottom graph represents slower fluctuations ($\tau=100$). In each case, we collected 100 samples, calculated temporal stability, and displayed all data points. We observed a qualitative change in the diversity–stability relationship depending on the pace of environmental fluctuations. In settings with fast environmental fluctuations, diversity and stability were positively correlated, whereas a negative correlation was observed in environments with slower fluctuations.
}
\label{diversity_stability}
\end{figure}

We also delved into the relationship between diversity and stability in dynamic environments.
In this study, we focused on the temporal stability \cite{lehman2000biodiversity, van2021unifying} of total biomass and how it relates to the species richness within our model. To manipulate diversity, we set the speciation rate $\nu$ to zero and varied the initial number of species $S$ in our simulations. Temporal stability was quantified as the reciprocal of the coefficient of variation of total biomass, i.e., the ratio of the mean to the standard deviation. Figure \ref{diversity_stability} elucidates the relationship between species diversity and the temporal stability of total biomass within our model. The upper graph corresponds to relatively rapid environmental fluctuations with $\tau=10$, and the lower graph represents slower fluctuations with $\tau=100$. For each setting, we calculated temporal stability taking 100 samples and displayed all data points. The parameters used in our analysis were $I=1.0 and L=20$, and the simulation time was $t_f=1000$ for each sample.

Our analysis revealed a qualitative shift in the diversity–stability relationship contingent upon the pace of environmental fluctuations. Under conditions marked by fast environmental changes, we observed a positive correlation between diversity and stability, which suggests that a higher variety of species enhances ecological stability under such situations. Conversely, we observed a negative correlation between diversity and stability in environments characterized by slower fluctuations, which means that species diversity loses the temporal stability of the ecosystem in less dynamic settings. 

\section*{The Resource Competition Model}
The second model used in our study is a differential equation model for resource competition.
The model is a variation of MacArthur's resource competition model \cite{macarthur1969species}, which encompasses the dynamics of resources, consumers, environmental fluctuations, and extinctions of species. In this model, the dynamics of three species competing for two types of resources are explored within a dynamic environment. Denoting the population of species $i$ at time $t$ as $n_i$, then $n_i$ is governed by the following equation:
\begin{align}
    \frac{dn_i}{dt} = \left(\sum_{j=1,2} a_{ij}R_j - c\right)n_i, \label{model1}    
\end{align}
where $R_j$ represents the amount of resource $j$ at time $t$, and $a_{ij}$ denotes the dependency proportion of species $i$ on resource $j$, with the constraint that $\sum_j a_{ij} = 1$. The parameter $c$ represents the total amount of resources required to sustain one individual consumer.

In line with MacArthur’s ecological theory \cite{macarthur1969species}, we consider resources themselves as biological communities. The dynamics of resource quantities $R_j$ are described by the following differential equation:
\begin{align}
    \frac{d}{dt} R_j = r_j\left(1-\frac{R_j}{K_j}\right)R_j - \sum_{i=1,2,3} a_{i,j}n_i R_j .
\end{align}
The first term represents the logistic growth of the resource, and the second term describes the decrease in the resource due to predation by consumers. Here, $K_j$ represents carrying capacities, and $r_j$ are intrinsic growth rates. By assuming that resource dynamics occur much faster than species dynamics, we can approximate the values of $R_j$ with the values at the steady state \cite{murray2002mathematical}. We then obtain the following equation:
\begin{align}
    R_j = K_j\left(1-\frac{1}{r_j}\sum_{i=1,2,3}a_{ij}n_i\right).    
\end{align}

To investigate ecosystem responses to environmental change, we introduce three elements.
First, we incorporate periodic oscillations of the carrying capacities, $K_j$, as a form of environmental variation \cite{nisbet1976population, legovic1984harvesting, fan1998optimal}. Resources 1 and 2 undergo alternating high and low growth periods with a fixed duration $T$. That is,
\begin{align}
    &\begin{bmatrix}
        K_1 \\
        K_2
    \end{bmatrix}
     = 
    \begin{bmatrix}
        K_0(1+ \alpha) \\
        K_0(1- \alpha)   
    \end{bmatrix} \quad \text{for}\,\, t \in [2mT,(2m+1)T]\\
    &\begin{bmatrix}
        K_1 \\
        K_2
    \end{bmatrix}
     = 
    \begin{bmatrix}
        K_0(1- \alpha) \\
        K_0(1+ \alpha)   
    \end{bmatrix} \quad \text{for}\,\, t \in [(2m-1)T,2mT],  \label{K12}
\end{align}
where $m$ is an integer, and the parameter $\alpha\in [0,1]$ controls the degree of environmental variability.

Second, each species is extinct when its population falls below a certain threshold value, $n_{\text{ex}}$. In natural ecosystems, populations often experience rapid declines below certain thresholds due to factors such as demographic stochasticity and the Allee effect \cite{gilpin1986minimum, shaffer1981minimum}. Implementing a simple truncation of population sizes to zero when they fall below a certain threshold is a simple way to capture these extinction processes.

Finally, specific resource utilization strategies are assigned to each species. These strategies are specified using a matrix:
\begin{align}
    A = \begin{bmatrix}
    a_{11} & a_{12} \\
    a_{21} & a_{22} \\
    a_{31} & a_{32}
    \end{bmatrix}
    =
    \begin{bmatrix}
        (1+\lambda)/2 & (1-\lambda)/2 \\
        (1-\lambda)/2 & (1+\lambda)/2 \\
        1/2 & 1/2
    \end{bmatrix} .\label{a}
\end{align}
The parameter $\lambda \in [0,1]$ quantifies the degree of specialization of species 1 and 2 to resources 1 and 2, respectively. When $\lambda = 1$, species 1 is a specialist that exclusively uses resource 1, species 2 is a specialist that exclusively uses resource 2, and species 3 is a generalist that evenly exploits both resources. As $\lambda$ decreases, the degree of specialization of species 1 and 2 diminishes; when $\lambda = 0$, all species become generalists. This comprehensive model integrates various components to explore how ecosystems respond to environmental fluctuations and encompasses resource–consumer interactions, extinctions, and strategy diversity among species.

\section*{Results Obtained under  the Resource Competition Model}
\subsection*{Phase Diagram}
\begin{figure}
\centering
\includegraphics[width=8.6cm]{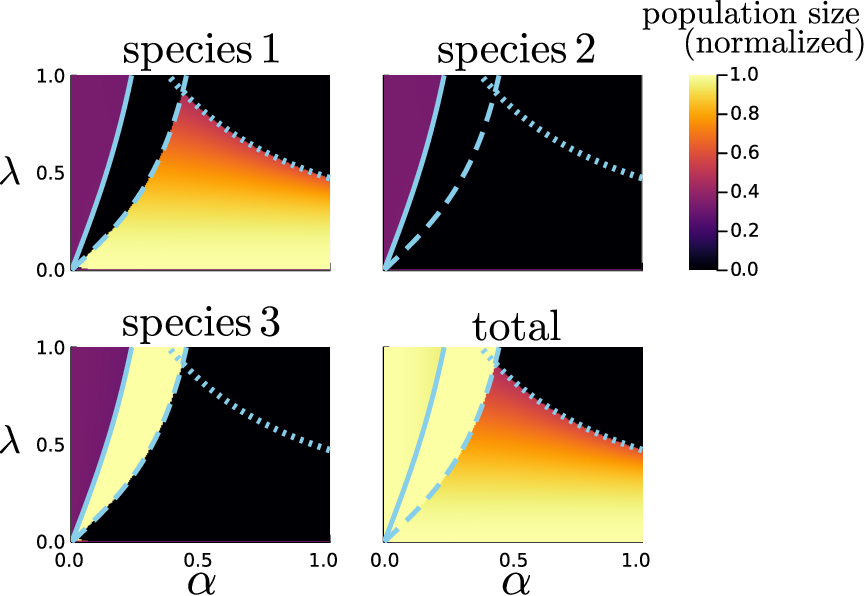}
\caption{
Simulation Results and Analytically Derived Phase Boundaries. This figure, which displays the results of simulations under the resource competition model, showcases the impact of environmental fluctuations ($\alpha$) and specialization ($\lambda$) on species dynamics. The cyan lines represent analytically derived phase boundaries that demarcate different ecological scenarios.
}
\label{plt4}
\end{figure}

Figure \ref{plt4} illustrates the results of simulations using the resource competition model. 
The horizontal axis represents the magnitude of environmental fluctuations, $\alpha$, and the vertical axis represents the degree of specialization, $\lambda$. We used parameters $c = 10$, $n_{\textrm{ex}} = 10$, $K_0 = 20$, $r_1 = r_2 = 50$, and $T = 1000$, which corresponds to a scenario in which environmental fluctuations are sufficiently slow compared with population dynamics. After setting initial population sizes to $n_1(0) = n_2(0) = n_3(0) = 50$, we calculated time-averaged population sizes from $t = 5T$ to $t = 10T$. By systematically varying the parameters $\alpha$ and $\lambda$, we obtained the heatmaps shown in Fig. \ref{plt4}.

The cyan lines in Fig. \ref{plt4} represent analytically derived phase boundaries, which perfectly match the simulation results.
These lines demarcate different ecological scenarios. Solid cyan lines correspond to conditions in a three-species coexistence state where species 2 becomes extinct. Dashed cyan lines correspond to situations in a coexistence state of species 1 and 3 in which species 3 becomes extinct. Dotted cyan lines represent conditions in a state with only species 1 present where species 1 becomes extinct. The simulation-derived phase boundaries closely align with the analytically obtained phase boundaries. For details of our analytical approach, please refer to the Materials and Methods section.

\begin{figure}
\centering
\includegraphics[width=7.5cm]{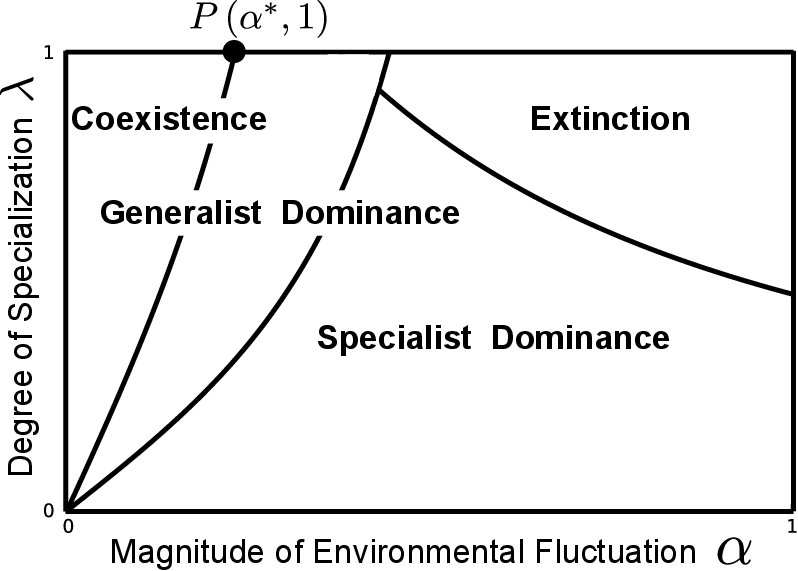}
\caption{
Phase Diagram of the Resource Competition Model. This diagram provides a comprehensive overview of the system’s behavior and illustrates the interplay between environmental fluctuations and the degree of species specialization. Four distinct phases, namely, Coexistence, Specialist Dominance, Generalist Dominance, and Extinction, depict different ecological scenarios within the model. The critical point $P(\alpha^*, 1)$ represents the threshold beyond which coexistence can no longer be sustained.
}
\label{Phase_Diagram}
\end{figure}

As shown in Fig. \ref{Phase_Diagram}, this system exhibits four distinctive phases, each reflecting a different ecological scenario.
In the Coexistence Phase, all three species are able to coexist without any species going extinct. This phase represents a state of balance where species 1, 2, and 3 share and use available resources effectively. The Specialist Dominance Phase is characterized by the dominance of specialist species, resulting in the extinction of the generalist. The selection of species 1 over species 2 within this phase is attributed to specific initial conditions, i.e., where resource 2 is scarce at the beginning of the simulation, as dictated by Eq. \ref{K12}. In the Generalist Dominance Phase, only the generalist species survives, and the other two specialist species go extinct. Finally, the Extinction Phase represents a scenario in which all species face extinction. This phase highlights the system’s sustainability when environmental conditions become particularly challenging.

\subsection*{Stability}
The phase diagram provides several pieces of information about the stability of the system.
First, we can observe the critical point $P(\alpha^*, 1)$, which is the point at which the three-species coexistence phase is disrupted. When environmental fluctuations remain moderate and $\alpha < \alpha^*$, the coexistence of three species is attainable. Beyond this threshold, however, three-species coexistence can no longer be sustained.

Second, specialization stabilizes the three-species coexistence state.
The phase boundary for three-species coexistence exhibits a positive slope. This result indicates that increased specialization enhances the system’s resilience to environmental fluctuations, thereby allowing the coexistence of all three species.

Our results also point to a potential trade-off: a heightened degree of specialization elevates the risk of overall ecosystem collapse. This trade-off is evident from the phase boundary for the Extinction Phase, which displays a negative slope. This situation suggests that the ecosystem’s susceptibility to complete extinction due to environmental variability intensifies as species 1 and 2 become more specialized.

\section*{Discussion}
In this study, we investigated the effects of escalating environmental fluctuations on ecosystems using two mathematical models: a stochastic ecosystem model with spatial structure, and a differential equation model for resource competition. In this section, we detail the key findings of our study and explore the underlying mechanisms as well as implications and future research prospects.

Our results clearly underscore the importance of specifying the characteristics of environmental fluctuations when assessing their effects on ecosystems. The period of fluctuations can dramatically alter the impact of a particular environmental stress on a given ecological community. Although multidimensional characterizations of ecosystem stability have previously been considered \cite{donohue2013dimensionality}, research that takes into account the diverse facets of environmental perturbations is still needed \cite{donohue2016navigating, ives2007stability}. By categorizing the type of perturbations and considering the unique features of the ecosystem under study, we can systematically investigate how disturbances influence ecological structure and diversity. The ability to predict a focal ecosystem’s robustness or vulnerability to specific environmental changes would provide valuable guidance for conservation and management efforts.

Qualitative shifts in the diversity–stability relationship of a system that are dependent on the period of environmental fluctuations can be comprehended by selecting specific adaptation strategies and sampling effects \cite{hector2002overyielding, huston1997hidden, tilman1997plant}.
Under rapid environmental fluctuations, generalists gain an advantage; with respect to the initial species composition, those taxa inclined toward generalist behavior exhibit a selective increase as well. As species diversity increases, the likelihood of sampling species with a stronger generalist tendency is heightened, resulting in an augmentation of temporal stability. In the face of slower-paced environmental fluctuations, in contrast, specialists tend to thrive. As the sample size grows, the likelihood of sampling highly specialized species is increased, in turn leading to a negative correlation between species diversity and temporal stability. To discern the subtle relationship between diversity and stability, the specific nature of environmental fluctuations must therefore be specified.

Our resource competition model offers a framework for comprehending the underlying mechanisms driving the coexistence, dominance, or extinction of specialist and generalist species. 
These processes, which unfold in four distinct phases, are intricately influenced by the degree of species specialization and the intensity of environmental fluctuations. The first phase, the Coexistence Phase, occurs in ecosystems characterized by moderate fluctuations and a certain degree of specialization. Under these conditions, both specialist and generalist species can coexist by carving out temporal niches \cite{huston1979general, harrington2009impact}. In the Generalist Dominance Phase, which emerges under conditions of higher environmental fluctuations, specialist species face extinction during resource scarcity due to their narrow niches. In contrast, generalist species can persist until the environment undergoes a shift, which ultimately results in generalist dominance. In the presence of even more drastic fluctuations, the Specialist Dominance Phase occurs. In this phase, generalist species are outcompeted by specialists, leading to the dominance of specialist species. A boundary exists, however, beyond which the Extinction Phase is realized. This phase occurs when highly specialized species initially outcompete other species but then become extinct owing to their vulnerability to environmental variability.

One critical insight from our study is the existence of a trade-off associated with specialization. Using the resource competition model, we investigated in detail how the degree of specialization impacts ecosystem stability. In this context, stability is the ability to remain in a particular state, which corresponds to ecological resilience \cite{holling1973resilience, gunderson2000ecological, standish2014resilience, scheffer2015generic, dakos2022ecological}. We observed that specialization enhances the stability of coexistence states. Specialization is a double-edged sword, however, as highly specialized ecosystems become more susceptible to total extinction when subjected to strong environmental fluctuations. This relationship can be viewed as an example of the trade-offs among multiple ecosystem functions \cite{zavaleta2010sustaining, raudsepp2010ecosystem, bennett2009understanding}. A future challenge is investigating whether this trade-off is a more general phenomenon that extends beyond our specific resource competition model.

In conclusion, our study has opened doors to a deeper understanding of how ecosystems respond to escalating environmental fluctuations. The knowledge gained serves as a foundation for future research aiming to explore the intricate relationship between environmental disturbances and their effects on ecosystems. By considering the specific characteristics of environmental fluctuations, we can improve our predictions and management strategies for different types of ecosystems. Furthermore, our findings point out the trade-offs associated with specialization, thereby highlighting the need for a careful approach to maintain biodiversity and ecosystem stability in an ever-changing world.

\begin{acknowledgments}
The author thanks Shin-ichi Sasa, Namiko Mitarai, and Ohta Hiroki for fruitful discussions. We thank Edanz for editing a draft of this manuscript. This work was supported by JSPS KAKENHI Grant No. 23KJ1335, a Grant-in-Aid for JSPS Fellows.
\end{acknowledgments}

\appendix \vspace{2cm}\hspace{2.9cm}
{\large \textbf{Appendix}}\\ \\
Here, we provide a detailed analysis of the resource competition model. Because we focused on situations where environmental fluctuations are sufficiently slower than population dynamics, we analyzed coexistence and extinction conditions by examining steady-state population equations.  We began by examining the conditions for the coexistence of all three species while considering steady-state equations and conservation law. Next, we investigated scenarios where two species coexist after one has gone extinct and derived explicit conditions for the extinction of the remaining species. Finally, we explored the case where only one species persists and determined the extinction conditions.

\subsection*{Coexistence of Three Species \label{3_analysis}}
We first consider the situation where all three species coexist.
The conditions for the steady state of Eq. \ref{model1} are
\begin{align}
A \textbf{R} = c \textbf{1} ,
\end{align}
where $\textbf{a} = [a_{11}, a_{12}, a_{13}]^\top, \textbf{1}= [1, 1, 1]^\top, A = [\textbf{a} \quad \textbf{1}-\textbf{a}]$, and $\textbf{R} = [R_1, R_2]^\top$. Except for the singular case where $\textbf{a} \parallel \textbf{1}$, $\text{rank}(A) = 2$. By defining the augmented coefficient matrix $\tilde{A}:= (A \quad c\textbf{1})$, we further find, except for the singular case, that $\text{rank}(\tilde{A}) = 2$. Because the rank of the coefficient matrix and the augmented coefficient matrix is equal to the number of unknowns, this system of linear equations has a unique solution: 
\begin{align}
    R_1 = R_2 = c. \label{3_ss_condition}
\end{align}

In addition, this system has a conserved quantity.
We rewrite Eq. \ref{model1} as
\begin{align}
    \frac{d}{dt} \left(\log{n_i}\right) = \sum_j a_{ij}R_j - c .
\end{align}
Taking linear combinations of these equations, we obtain
\begin{align}
    \frac{d}{dt} \left(\sum_i v_i \log{n_i}\right)
    = \textbf{v}^\top A \textbf{R} - c \textbf{v}^\top \textbf{1}, \label{vlogn}
\end{align}
where $\textbf{v}$ is a vector orthogonal to both $\textbf{a}$ and $\textbf{1}$:
    \begin{align}
    \textbf{v} = \textbf{a} \times \textbf{1} = 
    \begin{bmatrix}
        a_{21}-a_{31} \\
        a_{31}-a_{11} \\
        a_{11}-a_{21}
    \end{bmatrix}    .
\end{align}
With this choice of $\textbf{v}$, the right-hand side of Eq. \ref{vlogn} becomes zero. Consequently, the quantity
\begin{align}
    \gamma = \prod_i n_i^{v_i} \label{gamma} 
\end{align}
does not depend on time.
The conditions in Eq. \ref{3_ss_condition} and the conservation of $\gamma$ in Eq. \ref{gamma} determine the values of $n_1, n_2,$, and $n_3$ at the steady state.
To be more explicit, the values of $n_1$, $n_2$, and $n_3$ at the steady state are determined by the following three equations:
\begin{align}
    \sum_i a_{ij} n_i &= \mu_j \quad (j=1,2) \label{3_eq12} \\
    \prod_i n_i^{v_i} &= \gamma \label{3_eq3}, 
\end{align}
where
\begin{align}
\mu_j &= r_j \left(1 - \frac{c}{K_j}\right). \label{mu}
\end{align}

We next consider the condition for the extinction of species 1 in the three-species coexistence state.
At the moment of extinction, $n_1 = n_{\textrm{ex}}$, and Eq. \ref{3_eq12} thus becomes
\begin{align}
    a_{1j}n_{\textrm{ex}} + a_{2j}n_2 + a_{3j}n_3 = \mu_j \quad (j=1,2) .
\end{align}
Solving these equations for $n_2$ and $n_3$, we obtain 
\begin{align}
    \begin{bmatrix}
        n_2\\
        n_3
    \end{bmatrix}
 = \frac{1}{v_1}
    \begin{bmatrix}
     \:\:\,\, a_{32}\mu_1  - a_{31}\mu_2 + v_2 n_{\textrm{ex}} \\ 
        -a_{22}\mu_1 + a_{21}\mu_2 + v_3 n_{\textrm{ex}}
    \end{bmatrix}.
\end{align}
Substituting these expressions into Eq. \ref{3_eq3}, we obtain the extinction condition for species 1 in the three-species coexistence state:
\begin{multline}
    \left(\frac{a_{32}\mu_1 - a_{31}\mu_2 + v_2 n_{\textrm{ex}}}{v_1}\right)^{v_2}
    \left(\frac{-a_{22}\mu_1 + a_{21}\mu_2 + v_3 n_{\textrm{ex}}}{v_1}\right)^{v_3} \\
    = \, \gamma/{n_{\textrm{ex}}}^{v_1} \label{3_ex_condition}.
\end{multline}
Similar derivations can be performed for the extinction conditions of species 2 and 3.

\subsection*{Coexistence of Two Species}
We next consider the situations where two species coexist after the extinction of one species.
By following a similar argument as in the case of the three-species coexistence, the steady state condition of Eq. \ref{model1} is given by $R_1 = R_2 = c$, except for the singular case where species’ strategies are completely identical.

We consider the situation where species 2 has already gone extinct and look for the conditions under which species 3 goes extinct while species 1 and 3 coexist. At this moment, $n_2 = 0$ and $n_3 = n_{\textrm{ex}}$, and Eq. \ref{3_eq12} becomes
\begin{align}
    a_{11}n_1 + a_{31}n_{\textrm{ex}} &= \mu_1 \\    
    a_{12}n_1 + a_{32}n_{\textrm{ex}} &= \mu_2 . 
\end{align}
After eliminating $n_1$ from these equations, we obtain the condition for the extinction of species 3 in the two-species coexistence state:
\begin{align}
    a_{12}\mu_1 - a_{11}\mu_2 - v_2 n_{\textrm{ex}} = 0 \label{2_ex_condition}. 
\end{align}
Similar conditions can be derived for other scenarios.

\subsection*{Single-Species State}
Finally, we consider the state where only one species remains, and the other two have gone extinct. 
The condition for the steady state in this scenario is
\begin{align}
    \sum_j a_{ij}R_j = c ,
\end{align}
where species $i$ is assumed to be the last remaining species. Substituting the equation $R_j = K_j(1-a_{ij}n_i/r_j)$ into the above condition, we obtain the extinction condition for the last remaining species:
\begin{align}
    \sum_j a_{ij}K_j \left( 1-\frac{a_{ij}n_i}{r_j} \right) = c \label{1_ex_condition}.
\end{align}

\subsection*{Phase Boundaries}
Using these results, we plot the analytically derived phase boundaries on the $\alpha-\lambda$ plane.
We substitute Eqs. \ref{K12} and \ref{a} into the derived extinction conditions Eqs. \ref{3_ex_condition}, \ref{2_ex_condition}, and \ref{1_ex_condition}. The sign of Eq. \ref{K12} is chosen to be unfavorable for species that go extinct because we are focusing on the moment when the population first falls below the threshold $n_\text{ex}$. In this way, we obtain the analytical solution of the phase boundaries shown in cyan in Fig. \ref{plt4}.

\bibliography{ecosystem2023}
\end{document}